# The effect of hydrogen on the multiscale mechanical behaviour of a La(Fe,Mn,Si)$_{13}$-based magnetocaloric material

Siyang Wang[a,*], Oriol Gavalda-Diaz[a], Ting Luo[b], Liya Guo[a,c], Edmund Lovell[d], Neil Wilson[d], Baptiste Gault[b,a], Mary P. Ryan[a], Finn Giuliani[a]

[a]Department of Materials, Royal School of Mines, Imperial College London, London, SW7 2AZ, UK

[b]Max-Planck-Institut für Eisenforschung, Max-Planck-Straße 1, 40237 Düsseldorf, Germany

[c]School of Materials Science and Engineering, Shanghai University, Shanghai 200444, China

[d]Camfridge Ltd., Unit B1, Copley Hill Business Park, Cambridge Rd, Babraham, Cambridge, CB22 3GN, UK

*Corresponding author: siyang.wang15@imperial.ac.uk

**Keywords**: Magnetocalorics, Micromechanics, Hydrogen, Electron microscopy, Atom probe tomography

# Abstract

Magnetocaloric cooling offers the potential to improve the efficiency of refrigeration devices and hence cut the significant $CO_2$ emissions associated with cooling processes. A critical issue in deployment of this technology is the mechanical degradation of the magnetocaloric material during processing and operation, leading to limited service-life. The mechanical properties of hydrogenated La(Fe,Mn,Si)$_{13}$-based magnetocaloric material are studied using macroscale bending tests of polycrystalline specimens and *in situ* micropillar compression tests of single crystal specimens. The impact of hydrogenation on the mechanical properties are quantified. Understanding of the deformation/failure mechanisms is aided by characterization with transmission electron microscopy and atom probe tomography to reveal the arrangement of hydrogen atoms in the crystal lattice. Results indicate that the intrinsic strength of this material is ~3-6 GPa and is dependent on the crystal orientation. Single crystals under compressive load exhibit shearing along specific crystallographic planes. Hydrogen deteriorates the strength of La(Fe,Mn,Si)$_{13}$ through promotion of transgranular fracture. The weakening effect of hydrogen on single crystals is anisotropic; it is significant upon shearing *parallel* to the {111} crystallographic planes but is negligible when the shear plane is {001}-oriented. APT analysis suggests that this is associated with the close arrangement of hydrogen atoms on {222} planes.

# 1. Introduction

The realization of magnetic cooling technology relies on the availability and lifetime of high-performance active refrigerant materials – magnetocaloric materials. The magnetocaloric effect (MCE) is a magneto-thermodynamic phenomenon that results in the temperature change (heating or cooling) of a magnetocaloric material in response to a changing magnetic field. As a material it undergoes a magnetic ordering phase transition, causing it to heat up or cool down, provoked either by variation of temperature around the transition point (Curie temperature, $T_C$) or by the application and removal of external magnetic fields [1,2]. Magnetic refrigeration offers many environmental benefits including the reduction of $CO_2$ emissions (due to higher cooling efficiency) as well as avoiding the use of harmful refrigerants [3–7].

La(Fe,Si)$_{13}$-based intermetallics (face centred cubic crystal, NaZn$_{13}$-type structure) are relatively low-cost materials which exhibit a large MCE at the magnetic phase transition between the ferromagnetic and paramagnetic states. For compositions where the MCE is most pronounced the transition is of first-order, where the phase change in response to temperature, magnetic field or other external stimuli (such as pressure) is abrupt and accompanied by an abrupt change in lattice constant and therefore volume (although for this family the crystal symmetry is unchanged) [3,8–11]. Doping of the basic material (*e.g.* with substitutional Co, Mn and/or interstitial H) allows for tuning of $T_C$, close to which the MCE is largest, through alteration of the lattice parameter and exchange coupling [3,9]. H-doping (hydrogenation), in combination with some Mn substitution, is often considered the preferred route due to its retention of the favourable magnetocaloric properties. During processing, hydrogenation is the last step of the manufacturing process to tune $T_C$ (defined by the level of Mn-doping) to the desired temperature range for targeted applications, thereby bringing the operating temperature close to room temperature and enabling near-room temperature refrigeration.

A major challenge in utilizing La(Fe,Si)$_{13}$-based materials in a functioning cooling device is ensuring that the materials (and hence device) have sufficient lifetime. These materials are chemically reactive and have a low toughness [12,13]. Whilst the electrochemical behaviour of these materials in magnetic fields has been carefully studied and routes to protection developed [14], mechanical challenges still remain. The overall mechanical stability of these brittle materials is highly sensitive to small defects such as cracks. During manufacturing and operational processes, potential origins for crack development include mechanical load upon machining, volume expansion of the La-rich phase in air/water, thermal stress during hydrogenation, and volume change incompatibility at the first-order magnetic phase transition (the volume change is up to ~2% depending on the composition) [7,14–19]. Propagation of nucleated cracks through the material can immediately lead to failure of the entire refrigeration device.

Several approaches have been attempted to produce La(Fe,Si)$_{13}$-based materials with improved mechanical stability, through either embedding them into a soft matrix [20–23] or introducing an extra ductile phase in the La(Fe,Si)$_{13}$-based microstructure [24,25]. One of the simplest approaches is adding extra Fe into the material which leads to the formation of the ductile body centred cubic α-Fe phase that can act as barriers to crack propagation [12], and improves the overall strength of La(Fe,Si)$_{13}$-based magnetocaloric plates [24].

However, the in-service cracking issue, although considerably suppressed, has not been eliminated completely by addition of the α-Fe phase and is particularly severe for hydrogenated materials. This is consistent with mechanical testing results which showed a distinct decrease in strength post-hydrogenation [24]. The dominating mechanism for failure (i.e., is it the pre-existing cracks from machining and/or hydrogenation, the volume change upon magnetic transition, the magnetic field gradients, or other factors such as external mechanical loads due to the coolant flow), as well as the underlying mechanisms for hydrogen-induced degradation of mechanical performance remain largely unknown. Moreover, research studies on the mechanical performance of La(Fe,Si)$_{13}$ materials to date have not made it possible for the intrinsic properties of the La(Fe,Si)$_{13}$ compound to be directly measured [12,26–28], due to the pronounced effect of defects (such as cracks and pores) on mechanical property data and the difficulty of fabricating defect-free materials. Through instrumented nanoindentation, Glushko *et al*. [12] *estimated* the intrinsic strength of LaFe$_{11.2}$Si$_{1.8}$ to be ~2 GPa and also pointed out that micromechanical testing of *small* scale specimens is required not only to validate this estimation, but also to obtain information about the fundamental mechanical properties of La(Fe,Si)$_{13}$ compounds. Furthermore, hydrostatic stresses due to volume change in one crystal can result in shear stresses in the surrounding crystals, which may be the key origins for failure, as fatigue damage tends to accumulate due the build-up of plastic deformation which is a shear driven process. Small scale tests can be set up so the materials can fail in shear, which therefore enables studying the early stages of the plastic deformation that could be used as a basis for a more complex behaviour of the fatigue failure at a later stage of service life.

Understanding of the cyclic mechanical performance requires knowledge of a few fundamental aspects, such as how these materials deform, their strengths, and the effects of hydrogen. Such information of the intrinsic properties, as well as the mechanisms for hydrogen-induced failure, will enable the generation of models for lifetime prediction of La(Fe,Si)$_{13}$-based magnetocaloric materials under magnetic cycling, and inform the development of new material design with enhanced long-term structural integrity, essential for the successful realisation of important designs of cooling system based on this promising magnetocaloric material family.

Thus, in this work the focus is on the mechanical properties but within the context of the material's potential use for magnetocaloric-based cooling. The multiscale mechanical behaviour and properties of La(Fe,Mn,Si)$_{13}$ and the effects of interstitial hydrogen on them are investigated, using a combination of macroscale three-point bending tests of plate specimens and *in situ* compression tests of single crystal micropillar specimens inside a scanning electron microscope (SEM). Local deformation characters are examined by a transmission electron microscope (TEM) using thin foils lifted out from deformed pillars. In addition, atom probe tomography (APT) is employed to measure the hydrogen content in the specimen, and, crucially, to reveal the arrangement of hydrogen atoms which is shown to be linked to specific changes in the micromechanical properties.

## 2. Experiments

### 2.1. Material

Plates of a Mn-doped La(Fe,Mn,Si)$_{13}$-based magnetocaloric material were fabricated from bulk rods supplied by Vacuumschmelze GmbH & Co. KG. The material contains ~15% volume of α-Fe phase. The stoichiometry of the La(Fe,Mn,Si)$_{13}$ phase (referred to as 1:13 phase below), as measured by SEM-EDX (experimental details in section 2.3), is LaFe$_{12.4}$Mn$_{0.3}$Si$_{1.1}$. Some of the plates were (fully) hydrogenated through heat treatment in hydrogen atmosphere as described elsewhere [29]. For the hydrogenated plates, the temperature where the magnetic entropy change is maximum ($T_{peak}$) is ~15.6 °C, and so the materials used in the present work are in the paramagnetic state at room temperature (~20 °C) where the mechanical tests and electron microscopy characterization were carried out. APT experiments (details in section 2.4) were carried out on the hydrogenated material, in the ferromagnetic state, but we note that it has been shown previously that preferential habit sites for hydrogen (deuterium) in La(Fe,Si)$_{13}$ is independent of the magnetic state [30].

### 2.2. Mechanical testing

Three-point bending tests of the plate specimens were performed using a displacement-controlled Gatan Microtest 300 mechanical testing stage and a home-made bending fixture with a roller spacing of 8 mm. The loading speed used was ~0.2 mm·min$^{-1}$. Five samples of each type were tested.

For small scale tests of single crystal specimens, square micropillars with 2 μm mid-height width, 5 μm height, and 2° taper angle were fabricated using focused ion beam (FIB) on an FEI Helios Nanolab 600 dual-beam microscope. The FIB acceleration voltage was set to 30 kV, and a series of beam currents from 6.5 nA to 280 pA were used. High beam currents were employed to mill 20-μm-diameter circular trenches around the pillars for visualizing the deformation processes *in situ* during compression tests, while low beam currents were used for the final tailoring of the pillar contour. With the help of EBSD (experimental details in section 2.3), the micropillars were fabricated inside specific matrix (1:13 phase) grains with three groups of crystal orientations: <001>, <011>, and <111> directions of the crystals parallel to (±2° maximum) the out-of-plane direction (*i.e.*, the micropillar loading axis), respectively, to examine the mechanical response to loading along all lowest-index directions.

*In situ* micropillar compression tests were carried out using a displacement-controlled Alemnis nanoindenter in an FEI Quanta 650 SEM. The pillars were deformed with a 10 µm-diameter circular flat punch indenter supplied by Synton-MDP to achieve a (near) uniaxial stress state. The loading and unloading speeds employed were 5 nm·s$^{-1}$ and 50 nm·s$^{-1}$, respectively. Engineering stress was calculated by dividing the measured load with the mid-height cross sectional area of the pillar (as described in detail in Refs. [31,32]).

### 2.3. Electron microscopy characterization

EDX analysis was performed on a Zeiss Sigma 300 SEM with a Bruker XFlash 6-60 EDX detector at 20 kV beam acceleration voltage. This SEM was also used for imaging test pieces post-deformation.

For EBSD analysis, samples were mechanically ground with SiC abrasive paper and then polished with colloidal silica for 1 h. EBSD characterization was conducted on the FEI Quanta 650 SEM equipped with a Bruker eFlashHR (v2) EBSD camera, using a beam acceleration voltage of 30 kV and a probe current of ~10 nA. Collection and analysis of the EDX and EBSD data were carried out using Bruker ESPRIT 2.1 software.

To examine the local deformation features of the micropillars post-test, electron-transparent thin foil specimens for TEM analysis were prepared using a FIB in the FEI Helios Nanolab 600 dual-beam microscope. The specimens were imaged with a JEOL JEM-2100F TEM at 200 kV beam acceleration voltage.

### 2.4. Atom probe tomography

Needle-shaped APT specimens were prepared on an FEI Helios Nanolab 600 and a 600i dual-beam microscope, following the procedure described in Ref. [33]. Specific crystal orientations were targeted using EBSD prior to the FIB lift-out. In order to limit any H uptake from the chamber that may occur during the FIB milling process [34], and to ensure stability of the hydrogenated samples, the final annular milling steps were conducted at cryogenic temperatures down to approximately -190 °C. The cryo-stage setup and detailed cryogenic preparation steps are described in detail in Ref. [35]. APT measurements were performed using a commercial Cameca LEAP 5000XR atom probe microscope in laser pulsing mode at 50 K, with a pulse frequency of 125 kHz, laser energy of 15-20 pJ, and detection rate of 0.7-0.9%. APT data were then analysed using Cameca integrated visualization and analysis software (IVAS) integrated in AP suite 6.1.

## 3. Results

### 3.1. Microstructural characterization

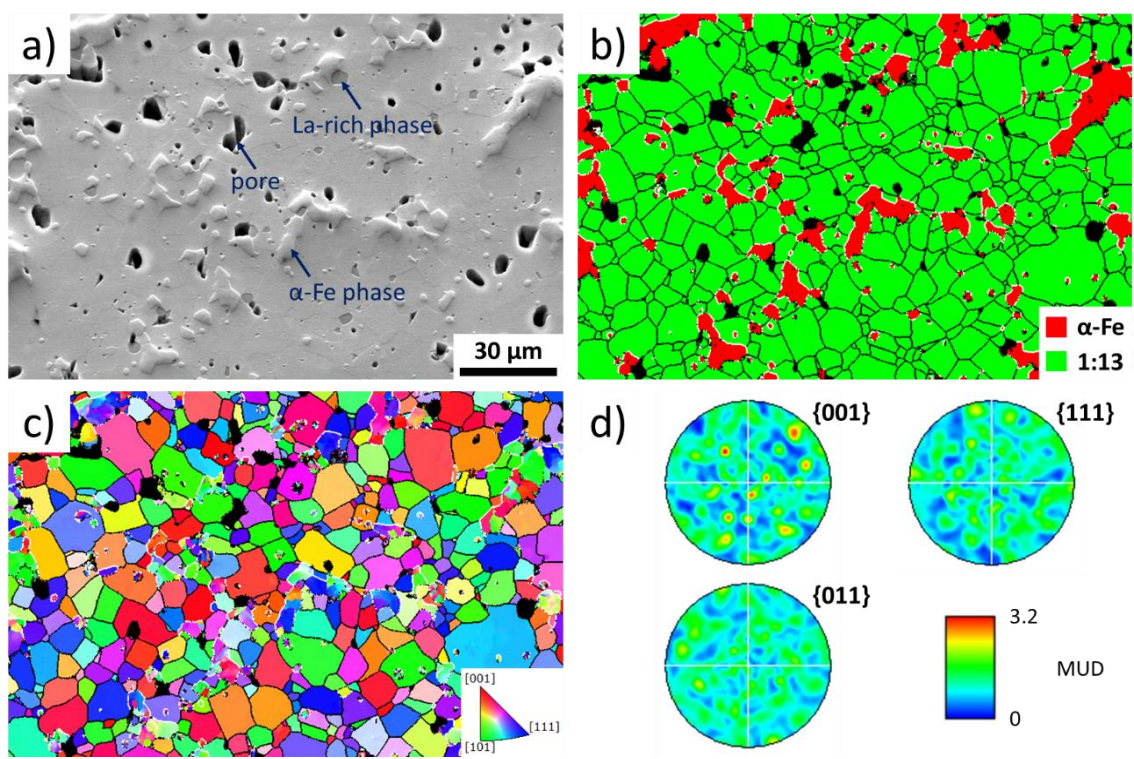

Figure 1 (a) Secondary electron image, (b) Phase map, (c) Inverse pole figure (IPF)-out-of-page map, and (d) pole figures (for the 1:13 phase) of an area on the as-received specimen. In (b) and (c), grain boundaries of the 1:13 phase, and phase boundaries between La(Fe,Mn,Si)$_{13}$ and α-Fe are highlighted with black and white lines, respectively. MUD denotes multiples of the uniform density.

The microstructure of the specimens, as characterized by electron backscatter diffraction (EBSD), is shown in Figure 1 where both 1:13 and α-Fe phases can be observed on the EBSD maps. The average grain size for the 1:13 phase is ~11 μm. The secondary electron image reveals the presence of a noticeable density of pores and some La-rich phase particles in the microstructure, which correspond to the non-indexed areas on the EBSD maps. According to Refs. [3,36], a certain degree of porosity can improve the mechanical stability of the material upon magnetic cycling, through relieving internal constraints at the grain boundaries and hence suppressing local stress build-up arisen from the volume change upon magnetic transition. The pole figures for the 1:13 phase indicate that the grains are randomly oriented with no texture present, in agreement with Ref. [15]. No microstructural difference was observed between the as-received and hydrogenated specimens.

## 3.2. Three-point bending tests

The results of the three-point bending tests are shown in Figure 2. The load-displacement responses for both type of specimens show characteristic brittle fracture behaviour, with load drop to zero point after linear elastic regime without any noticeable plastic flow (as per Ref. [37]). For the as-received (non-hydrogenated) specimens, the fracture strength and strain measured here (230 MPa and 0.22%) are in good agreement with three-point bending test results for $La_{1.7}Fe_{11.6}Si_{1.4}$ plates [37], and close to the lower bound values obtained through compression tests of $LaFe_{11.2}Si_{1.8}$ (180-620 MPa and 0.2-0.6%) [12]. The fracture strength and strain for the as-received specimens are both ~25% higher than those for the hydrogenated ones, while the two types of specimens show similar bending modulus values that agree well with those measured in Refs [12,26] at room temperature (~100 GPa).

SEM imaging of the fracture surfaces was performed on the fractured test pieces to aid in understanding of the observed differences in fracture strength and strain between as-received and hydrogenated specimens, and the results are shown in Figure 3. The fracture surfaces of the as-received specimens are rough in general. The 3D morphology of the grains can be observed, indicating that the fracture paths are mainly along grain boundaries (*i.e.*, intergranular fracture). Unlike the as-received specimens, the hydrogenated specimens show nearly completely flat fracture surfaces without intergranular fracture characters, indicating a change in the crack propagation behaviour that is mainly through the grain interior (*i.e.*, transgranular fracture).

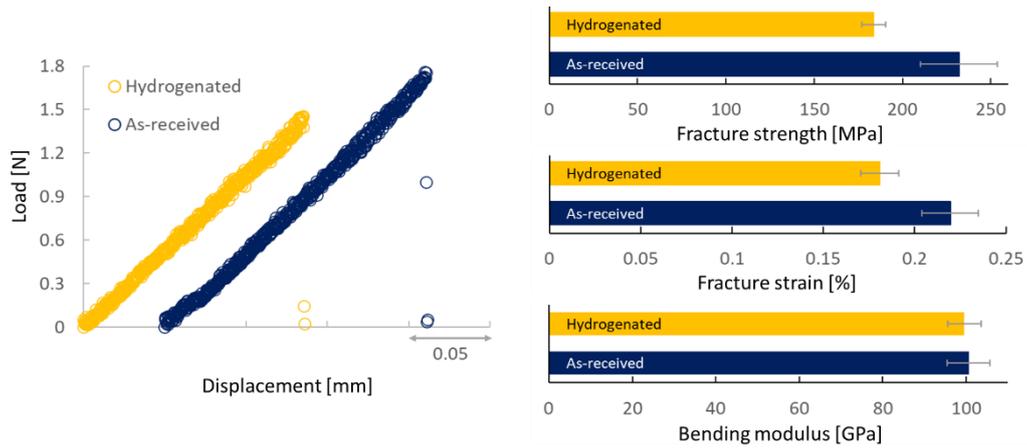

Figure 2  Typical load-displacement curves for the as-received (non-hydrogenated) and hydrogenated plate specimens recorded during the three-point bending tests, and fracture strength, fracture strain and bending modulus values averaged from five tests of each type of specimen. Error bars represent standard deviation.

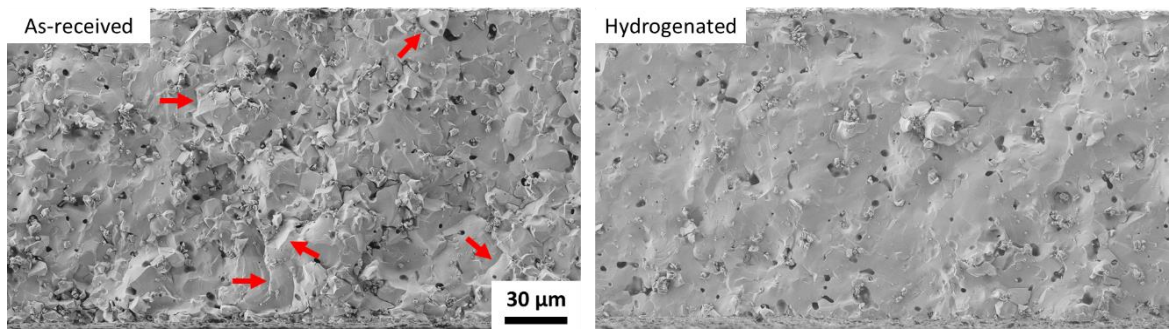

Figure 3  Secondary electron images of the fracture surfaces for the as-received and hydrogenated specimens after three-point bending tests. Grain features (such as the grain boundaries marked with the red arrows) can be observed on the fracture surface of the as-received sample, indicating intergranular fracture, as opposed to the transgranular fracture surface of the hydrogenated sample.

### 3.3. Micropillar compression tests

The engineering stress-indenter displacement responses of the micropillars in the as-received and hydrogenated specimens during compression tests are plotted in Figure 4, with each colour representing a certain group of crystal orientations. For both samples, two <001>- and <011>-oriented pillars were tested, and good consistency in failure stress is observed between pillars of each type (see the red and yellow curves in Figure 4, the two curves for each orientation nearly overlap prior to the stress drops). Only one <111>-oriented pillar for each sample was made and tested due to lack of large enough grains with this orientation.

We therefore complemented this through testing smaller pillars (1 × 1 × 2.5 μm$^3$) which allows the use of smaller grains. The stress-displacement responses obtained agree well with those for the larger pillars of the same type. Test results for the smaller pillars are given in Figure S1, Supplementary material. Further discussion on this is provided in section 4.1.

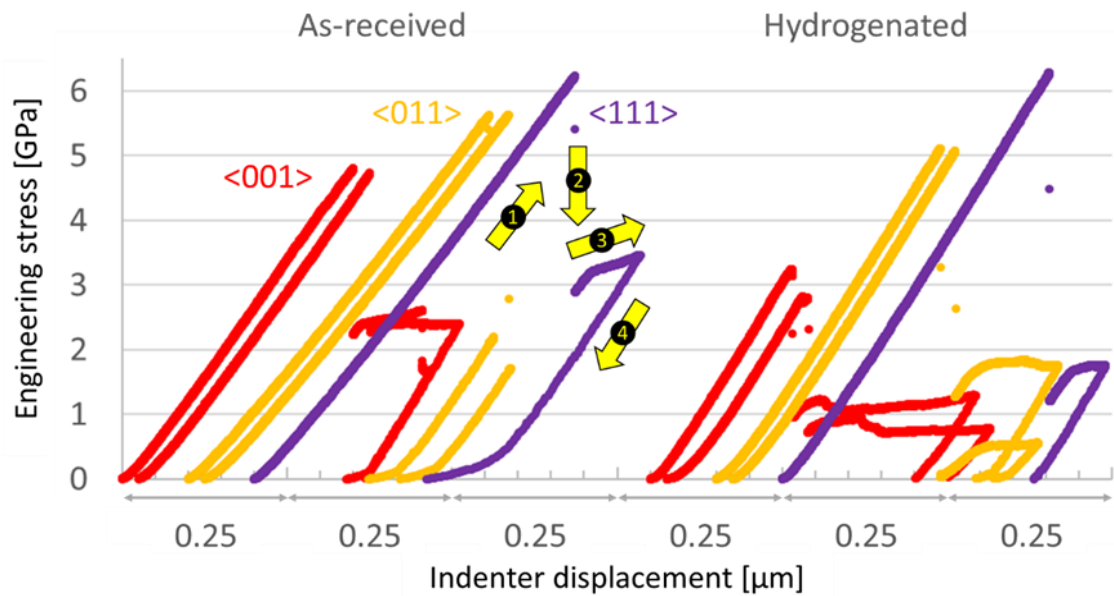

Figure 4  Engineering stress-indenter displacement curves for the 2 × 2 × 5 μm$^3$ micropillars in the as-received and hydrogenated specimens. Curves for micropillars with <001>, <011> and <111> crystallographic directions parallel to the loading axis are plotted in red, yellow, and purple, respectively. The 4 stages involved in the testing process of a micropillar are labelled on the curve for the "As-received, <111>-oriented" pillar: ①initial loading, ②stress drop, ③further loading, ④unloading.

For all the pillars tested, linear elastic deformation is followed by distinct stress drops on the stress-displacement curves with no observable plastic flow. Upon further loading, the engineering stress level for each pillar is relatively constant and not able to get back to the original stress level before the drop. For both the as-received and hydrogenated samples, the strengths are the highest for the <111>-oriented pillars and the lowest for the <001>-oriented pillars. The strengths of all the pillars tested are summarized in Table 1.

Comparison is made between the as-received and hydrogenated pillars for each group of crystal orientation. For pillars with <001> and <011> directions parallel to the loading axis, the as-received pillars are stronger than the hydrogenated pillars. For the <111>-oriented pillars, on the other hand, the strengths of the as-received and hydrogenated pillars are of comparable values.

Table 1 Shear planes and strengths for the micropillars tested. Errors come from tapering of the pillars and noise in load measurement.

| Loading axis | | <001> | | <011> | | <111> | |
|---|---|---|---|---|---|---|---|
| Pillar size [μm³] | | 2 × 2 × 5 | | 2 × 2 × 5 | | 2 × 2 × 5 | 1 × 1 × 2.5 |
| Shear plane | | {111} | | {111} | | {001} | |
| Sample | | #<001>-1 | #<001>-2 | #<011>-1 | #<011>-2 | #<111>-1 | #<111>-2 |
| Strength [GPa] | As-received | 4.8 ± 0.8 | 4.7 ± 0.8 | 5.6 ± 1.0 | 5.6 ± 1.0 | 6.2 ± 1.1 | 6.2 ± 1.1 |
| | Hydrogenated | 2.8 ± 0.5 | 3.2 ± 0.6 | 5.1 ± 0.9 | 5.1 ± 0.9 | 6.3 ± 1.1 | 6.7 ± 1.2 |

Figure 5 presents the post-deformation SEM images of the micropillars in both specimens and with different crystal orientations. It is evident that all the pillars shown exhibit highly similar mechanical behaviours where the applied strain is accommodated by the formation of a localized shear band on each pillar. Note that we use the term "shear band" here, as there is ambiguity in the formation mechanism of these deformation features (as will be discussed in section 4.1). This should be distinguished from slip bands which are normally observed in materials that deform plastically *via* dislocation motion. During the *in situ* SEM compression tests, it was observed that for each test the shear band formed simultaneously with the stress drop on the stress-displacement response. It is also worth noting that 1 out of the 12 pillars tested was found to had broken into two pieces after the indenter tip was retracted (Figure S2, Supplementary material).

Based on the knowledge of the crystal orientations of the pillars from EBSD, the Miller indices of the crystallographic planes which the shear planes are parallel to were extracted. The shear planes for the <001>- and <011>-oriented pillars are both {111}-oriented. The <111>-oriented pillars for which the resolved shear stresses on {111} planes are (near-)zero upon uniaxial compression, show shear planes parallel to the {001} planes. Therefore, the as-received and hydrogenated pillars exhibit a more dramatic difference in strength upon shearing parallel to the {111} planes (*i.e.,* the <001>- and <011>-oriented pillars), than to the {001} planes (the <111>-oriented pillars, see Table 1).

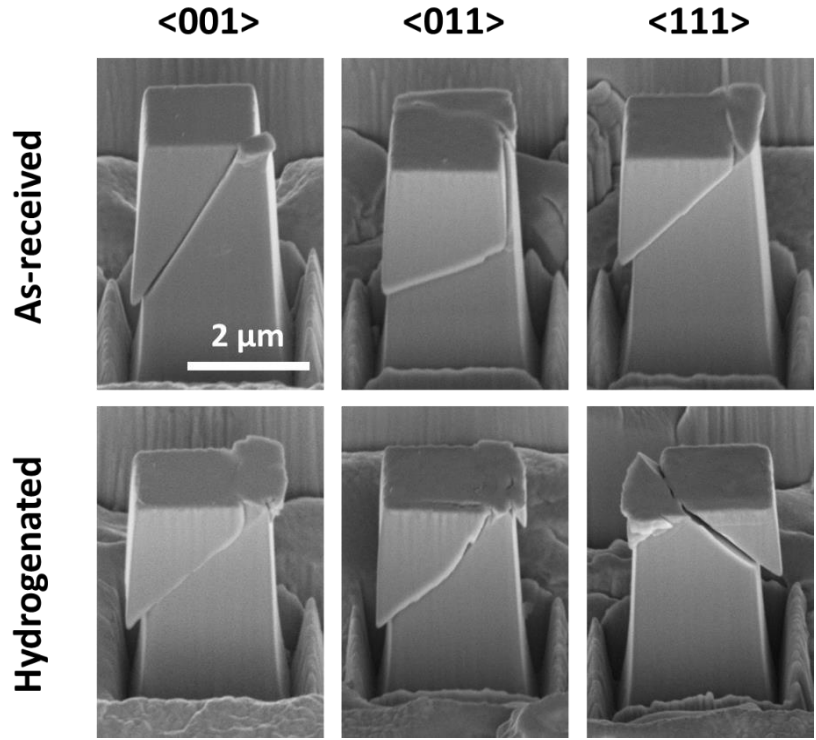

Figure 5 Post-deformation SEM images of the micropillars in the as-received and hydrogenated specimens. The Miller indices indicate the loading axis in the crystal frame.

The shear directions for pillars with {111}- and {001}-oriented shear planes are parallel to <112> and <011> crystallographic directions, respectively. The critical resolved shear stresses (CRSS, $\tau_c$) for each shear mode were then calculated based on the crystal orientations and the strengths of the pillars tested. For as-received material, the CRSS is 2.5 ± 0.7 GPa for {111} <112>-type shear, and 2.9 ± 0.5 GPa for {001} <011>-type shear, which change into 2.0 ± 0.9 and 3.1 ± 0.6 GPa respectively for hydrogenated material.

Bright-field TEM images of the shear band formed on a <011>-oriented micropillar in the as-received specimen are given in Figure 6. The pillar from which the TEM sample was lifted out was unloaded immediately after shear band formation (*i.e.*, at the stress drop point on the stress-displacement response) during the compression test, in order to avoid any effect of further loading on the local structure around the shear band. The image shows that for this pillar, the localized shear band observed in the SEM comprises two parallel shear bands closely spaced (~100 nm) to each other (as highlighted with the yellow dashed lines). Some dislocations exist at and near the shear bands, as marked with the red arrows in Figure 6(a). A dislocation loop (the leftmost marked structure on the image) can be observed at the shear band on the top. Additionally, there are dislocations seemingly bowing out from the lines between pinning points (marked with the black arrows). Figure 6(b) shows a high-magnification view of the region within the black box in (a). No clear indication of dislocations is observed here, while the high signal intensity along the shear band implies the formation of two new surfaces likely through fracture.

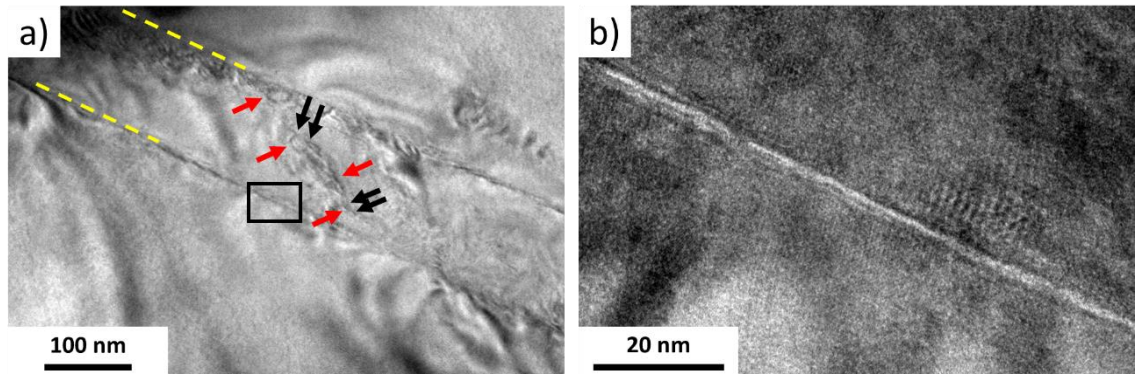

Figure 6 (a) Bright-field TEM image of the shear band formed on a <011>-oriented micropillar in the as-received specimen. The shear bands are highlighted with the yellow dashed lines. Dislocations are marked with the red arrows and pinning points with black arrows; (b) high-magnification view of the region within the black box in (a).

## 3.4. Atom probe tomography

The chemical compositions of the 1:13 phase in the as-received and the hydrogenated samples were determined using APT, and the results are shown in Table 2. The La:Fe:Si:Mn ratio obtained agrees with SEM-based energy-dispersive X-ray spectroscopy (EDX) measurements (section 2.1). The 1:13 phase in the hydrogenated sample contains 13.1 ± 1.5 at.% of H, which is close to the reported deuterium concentration in fully hydrogenated La(Fe$_{0.88}$Si$_{0.12}$)$_{13}$ (~10%) obtained from Rietveld analysis of neutron diffraction patterns at 350 K [30]. The H concentration in the 1:13 phase for the hydrogenated sample is over twice that for the as-received sample which contains 5.2 ± 0.8 at.% of H. Unlike the matrix phase, the amount of H in the α-Fe phase are similar for both samples: 3.8 ± 0.007 at.% *vs*. 2.9 ± 0.008 at.% for the as-received sample and the hydrogenated sample respectively, as shown in Table S1 and S2, Supplementary material.

Such a high level of H content in the as-received sample can be in part from the material itself, in part from the preparation and/or from the residual gases in the ultra-high vacuum chamber of the atom probe. Further discussions on these aspects can be found in Ref. [38] for instance, it is well established that the detection of spurious hydrogen is related to the strength of the electrostatic field applied during the analysis [39], which can be traced back by the ratio of charge-states [40–43]. Across the four datasets analysed for each of the material's states, the average $Fe^{2+}/(Fe^{1+}+Fe^{2+})$ ratio was 99.96±0.002% and 99.78±0.05% for the as-received and hydrogenated samples respectively. This indicates comparable electrostatic field conditions, and therefore the change in H concentration in the 1:13 phase can be ascribed to the hydrogen-charging process, not to an artefact from APT.

Table 2 Chemical composition (at. %) of the 1:13 phase in the as-received and the hydrogenated specimens, as measured by APT.

|  | Fe | La | Si | Mn | H |
| --- | --- | --- | --- | --- | --- |
| As-received | 76.19 ± 0.61 | 7.43 ± 0.11 | 8.88 ± 0.07 | 2.11 ± 0.01 | 5.16 ± 0.78 |
| Hydrogenated | 19.83 ± 1.20 | 6.77 ± 0.17 | 8.21 ± 0.18 | 1.89 ± 0.03 | 13.06 ±1.55 |

Information from EBSD was used to guide the preparation of specimens for APT analysis along specific crystallographic directions in order to exploit the optimal spatial resolution in the depth of the specimen [44,45]. This allows us to reveal the location of hydrogen at the atomic level. Different poles are observed in the APT volumes through the 2D density maps, enabling the exploitation of the retained crystallographic information [46], in particular *via* the use of species-specific spatial distribution maps (SDM) [47]. Regions of interest (ROI) were centred on the poles so as to resolve individual sets of atomic planes. SDMs were then calculated along the z-direction thereby revealing the average distance between Fe and H atoms, which can be used to study site occupancy in ordered structures [48,49].

APT volumes were obtained for grains oriented near the [111] and [110] directions in the as-received sample. It was observed that only *z*-SDMs for Fe-Fe exhibit a set of peaks with spacing corresponding to that for the {222} and {220} planes following adjustment of the reconstruction parameters as described in [50]. Figure S3, Supplementary material shows the APT data for the [111]-oriented grain. No specific site occupancy is observed in the *z*-SDMs for H-H. This contrasts with the hydrogenated sample where APT analyses were performed in grains oriented near the [111] and [100] zone axes. The 2D ion density maps and the resolved atomic planes are shown in Figure 7(a) and (b). The associated *z*-SDMs in Figure 7(c) show that Fe and H exhibit the same site occupancy with an average spacing of 0.33 nm, which corresponds to the interplanar distance of the {222} planes. Along the [100] direction, periodic peaks with an average spacing of 0.57 nm are found in the *z*-SDM of Fe (Figure 7(d)). H atoms however appear nearly *randomly* distributed, with no clear peak observed in the *z*-SDM of H-H. Yet the compositions measured for these two grains are very close: 13.36 at.% of H for the [111]-oriented grain vs. 14.44 at.% of H for the [001]-oriented grain. The difference in the SDMs therefore cannot be ascribed to a specific loss of H during the analysis that could obscure the ordering in the SDMs.

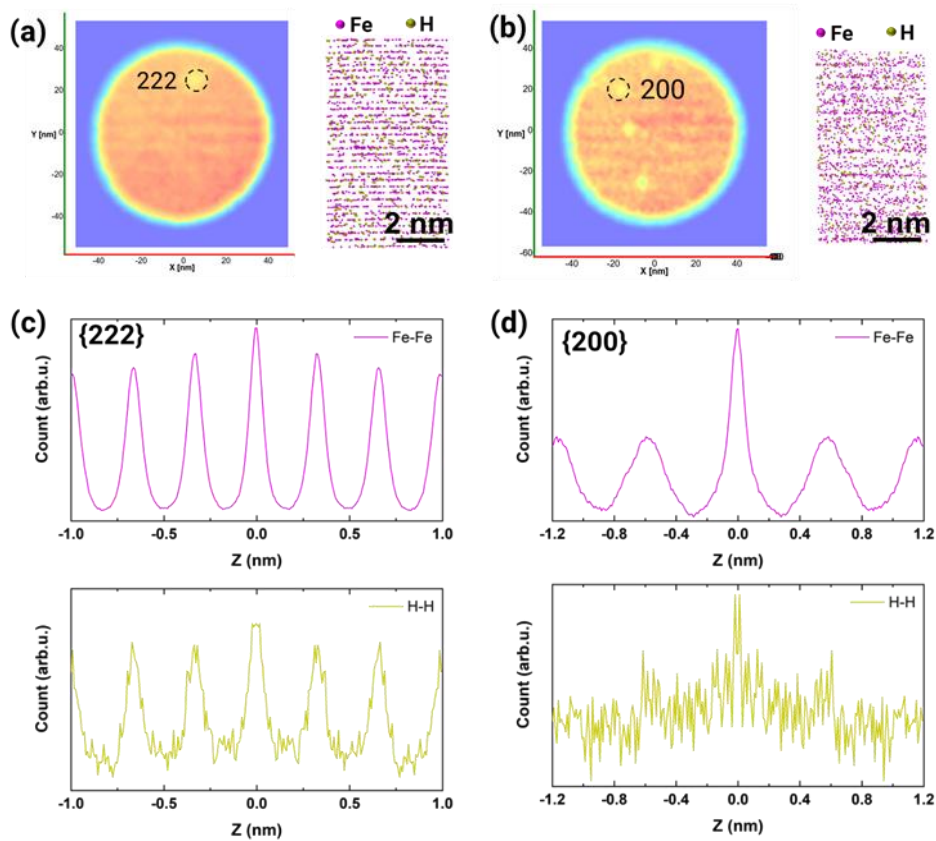

Figure 7 APT analysis of the hydrogenated sample: 2D density map of the APT volume taken from two grains oriented near (a) [111] and (b) [100] zone axes. ROIs are centred within the poles and the resolved {222} and {200} planes are shown next to the 2D density maps. (c), (d) *z*-spatial distribution maps of Fe and H corresponding to the {222} planes in (a), and the {200} planes in (b), which clearly show ordered arrangement of H on {222} planes and random distribution on {200} planes. Arb.u. denotes arbitrary unit.

# 4. Discussion

## 4.1. Intrinsic mechanical properties of La(Fe,Mn,Si)$_{13}$

Under load, the active (low-index) shear planes for La(Fe,Mn,Si)$_{13}$ are parallel to the {111} and {001} crystallographic planes, depending on the crystal orientation with respect to the loading configuration. The intrinsic strengths of the material tested (Table 1) are 4.8 ± 0.8, 5.6 ± 1.0 and 6.2 ± 1.1 GPa for the <001>-, <011>- and <111>-oriented crystals, respectively, while hydrogenation brings the strengths down to 3.0 ± 0.6 and 5.1 ± 0.9 GPa for <001>- and <011>-oriented pillars respectively. There was little change for the <111> orientation [6.2 ± 1.1 *vs*. 6.5 ± 1.2 GPa]. Broadly, the strength values obtained in this work indicate that the estimation by Glushko *et al*. [12] (~2 GPa) is an underestimate but is of the correct order.

The significant stress drops observed on the stress-displacement curves for the pillar compression tests (akin to the strain bursts in load-controlled tests) could be an indicator of either dislocation nucleation or the generation of cracks. However, it is hard to differentiate between these two for the material tested here. Although dislocation structures have been observed on the TEM images, the general appearances of the pillars in the SEM as well as the magnified view of the shear band in the TEM imply that the samples likely failed in a brittle manner. This is rare in small scale tests where shear fracture is not a commonly-observed fracture mode [51]. In fact, the complicated structure of this material, and the large size of its unit cell (Figure 8) suggest that dislocation movement in the crystal might be fairly difficult due to long Burgers vectors. Detailed interactions between plasticity and fracture processes, particularly which of the two initiates first at the point where the stress drops took place, is intractable based on the results obtained so far. Further development in clarifying these phenomena could be made possible through *in situ* TEM experiments or computational simulation approaches. This would shed light on the fatigue behaviour of these materials, which is crucial for understanding in-service failure under magnetic field cycling.

It is known that for (macroscopically) brittle materials, plasticity may be favoured over fracture when test pieces are smaller than critical sizes [52,53], and therefore plasticity and fracture features might co-exist when the size is at the boundary between the two regimes. However, the fact that reducing the pillar volume by a factor of 8 did not change the deformation behaviour of La(Fe,Mn,Si)$_{13}$ pillars (Figure S1, Supplementary material) means that the observed phenomena are not likely due to the size effect, but are intrinsic properties of the material itself. In addition, the strengths of the <111>-oriented pillars for each type were also found to be size-insensitive (Table 1), which once again confirms that this is not a size effect for the pillar dimensions employed.

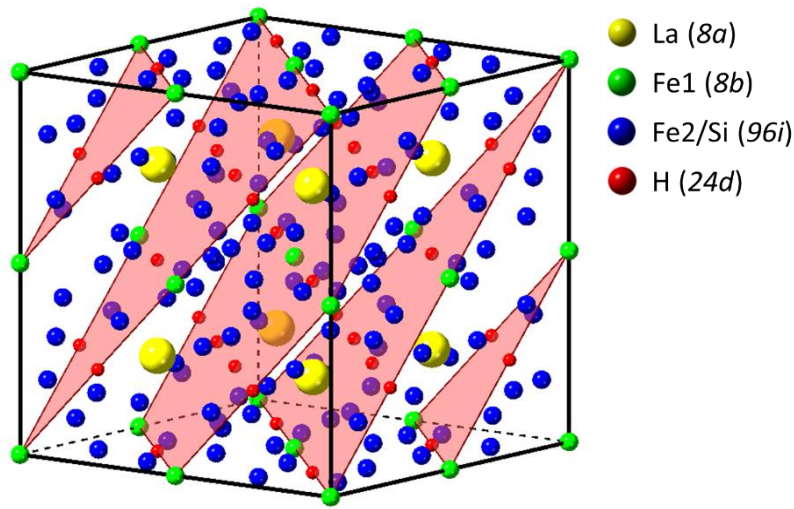

Figure 8  The crystal structure of La(Fe,Si)$_{13}$H$_x$, generated according to Ref. [54]. The {222} planes are shaded. X-ray diffraction (XRD) pattern of the hydrogenated material studied can be found in Figure S4, Supplementary material.

## 4.2. Effect of hydrogen on the mechanical behaviour of La(Fe,Mn,Si)$_{13}$

The results of the three-point bending tests have shown that hydrogen deteriorates the strength of La(Fe,Mn,Si)$_{13}$, and fractography revealed that there is a switch in fracture path from being mainly intergranular to nearly completely transgranular after hydrogenation. This suggests that the crystal lattices of La(Fe,Mn,Si)$_{13}$ have been weakened by the insertion of interstitial hydrogen atoms. This was confirmed by micropillar compression tests of single crystal specimens where the hydrogenated pillars exhibit lower strength than the as-received pillars, particularly upon shearing parallel to {111} planes. The change in crack path observed also indicates that the weakening effect of hydrogen on grain boundaries is not as significant as it is on crystal lattices for La(Fe,Mn,Si)$_{13}$, which is consistent with the high solubility of hydrogen in this material.  In materials with relatively low hydrogen solubility (such as steels) the fracture path is often promoted *towards* the intergranular direction by the addition of hydrogen atoms which mostly reside at grain boundaries [55].

The shear plane-dependent weakening effect of hydrogen on the strengths of the micropillars is found to be correlated to the arrangement of hydrogen atoms on crystallographic planes as revealed by APT: a pronounced effect of hydrogen on pillar strength is observed when the shear plane is {111}-oriented, which interestingly corresponds to alignment of hydrogen atoms in the crystal (see the SDM of H for the {222} planes in Figure 7). For the {002} planes on which the arrangement of hydrogen atoms is random, there is no significant effect of hydrogen on pillar strength when the shear planes are on the same orientation. Our APT results support prior work in the literature which reported (according to neutron diffraction data) that hydrogen atoms in La(Fe,Si)$_{13}$H$_x$ occupy the 24*d* sites of the crystal lattice [30,56], which would lead to the closest arrangement of hydrogen atoms on {222} planes as highlighted in Figure 8, while the hydrogen atoms do not explicitly habit on {002} planes. Together these suggest that hydrogen deteriorates the strength of La(Fe,Mn,Si)$_{13}$ crystal *anisotropically*, with the weakening effect being significant on {222} crystallographic planes where hydrogen atoms are arranged closely, but negligible on {002} planes where the areal density of hydrogen atoms is low.

As mentioned previously, hydrogen is added into the lattice in order to tune the $T_C$ of the material. Our micromechanical tests indicate that this results in a detrimental effect on the mechanical properties through weakening specific crystallographic planes. Therefore improvements in the mechanical stability of La(Fe,Mn,Si)$_{13}$-based materials could be achieved not only *via* introducing ductile constituents, but also through fabrication of textured materials such that the anisotropic negative effect of hydrogen on the mechanical performance could be suppressed. This, along with the generation of simulation models using the intrinsic properties obtained in this work, will enable deeper understanding of the mechanical behaviour of these materials, and effective design of new material microstructures that provide long-term durability.

### 4.3. Implications for mechanical stability over magnetic cycling

From the test results we note that the failure strength and strain in single crystal micropillar compression are both ~20 times that in bending of polycrystal aggregates. This likely arises from three potential origins:

a. In bending tests, materials normally fail as a result of tensile stress due to its high sensitivity to defects, while strength obtained in compression tests are closer to the intrinsic properties of the material;
b. The possible defect size involved in micropillar compression tests is much smaller than the bending tests due to the small volume of the specimens, and the pillars were intentionally fabricated in defect-free regions of the materials;
c. The strain energy volumes are different for these two geometries.

While results from the bending tests reflect the strength of the material in use, micropillar compression tests provide the ideal strength level that could be achieved in a material with minimized defect population. We have shown here that this material is fundamentally strong, however the mechanical performance of the current generation of products may have been limited by the large number of cracks (as per Ref. [12]) that are present.

As introduced earlier, shear stresses (rather than hydrostatic stresses) are more likely the direct origin for fatigue, as fatigue damage tends to accumulate due the build-up of plastic deformation which is a shear driven process. The shear properties of the material obtained in this work would therefore provide important input for the simulation of fatigue failure. Furthermore, our results suggest that it is crucial to revisit the current knowledge about in-service mechanical degradation of $La(Fe,Si)_{13}$ where the transition-induced volumetric stress is seemingly accepted as the dominating mechanism, while contributions from the large number of pre-existing cracks, the magnetic field gradient, and other external stresses (such as pressure from the heat transfer fluid running through the channels) are often neglected. It is hence vital to examine fatigue damage for materials exposed to each of these potential sources individually, thereby evaluating their relative significance. Clarifying the origin for failure would also inform potential texture design, if the stress state for the main driver of crack introduction is anisotropic.

Incorporating $La(Fe,Mn,Si)_{13}$ into a working magnetocaloric device with reliable mechanical stability would be significantly more achievable if the failure of the material arises not from its intrinsic properties, but from artefacts during the manufacturing process. Attentions should be paid to quality control of the materials used, and research efforts into the durability-quality relationship are required to help solve the engineering problem.

## 5. Conclusions

The mechanical behaviour of a La(Fe,Mn,Si)$_{13}$-based material across a range of length scales and the effect of hydrogen have been investigated using three-point bending tests of polycrystalline plates and *in situ* SEM compression tests of single crystal micropillars combined with TEM and APT analyses. The following conclusions can be drawn.

1. The active deformation mode for single crystal La(Fe,Mn,Si)$_{13}$(H$_x$) under uniaxial compression is localized shearing *parallel* to the {111} and {001} crystallographic planes, and along <112> and <011> directions respectively.
2. For as-received material, the critical resolved shear stress is 2.5 ± 0.7 GPa for {111} <112>-type shear, and 2.9 ± 0.5 GPa for {001} <011>-type shear, which change into 2.0 ± 0.9 and 3.1 ± 0.6 GPa respectively for hydrogenated material.
3. The deformation processes of the micropillars exhibit a combination of plasticity and fracture characters, which is found to be specimen size-independent and is therefore an intrinsic behaviour of the material.
4. The size-independent strengths of the single crystal micropillars are ~3-6 GPa, with the <001>- and <111>-oriented pillars (with respect to the loading axis) being the weakest and strongest, respectively.
5. Hydrogen deteriorates the strengths of the micropillars for which the shear planes are {111}-oriented but has negligible effect on the strengths of the pillars that shear *parallel* to the {001} planes.
6. The anisotropic effect of hydrogen on pillar strength is related to the ordered arrangement of hydrogen atoms in the crystal lattice, where they stack closely on {222} planes but have low areal density on {002} planes as revealed by APT.
7. For polycrystalline plates under bending, this weakening effect of hydrogen on the crystal lattice of La(Fe,Mn,Si)$_{13}$ has led to a change from intergranular to transgranular fracture post-hydrogenation, as well as decreased strength.
8. La(Fe,Mn,Si)$_{13}$ is fundamentally strong, however the strength of the materials in use is limited by the large number of pre-existing cracks in the microstructure. It is therefore crucial to suppress crack formation during manufacturing and evaluate its significance for in-service mechanical stability.

# Acknowledgements

MPR, FG, NW, EL and SW acknowledge funding from Innovate UK (UKRI 32645). TL acknowledges the financial support from the Alexander von Humboldt Foundation. We thank Mr. Garry Stakalls for manufacturing the fixture used in the three-point bending tests, and Yiyang Cai for the XRD data in Supplementary material. The FEI Quanta SEM used was supported by the Imperial-Shell AIMS UTC and is housed in the Harvey Flower EM suite at Imperial College London.

**Supplementary material**

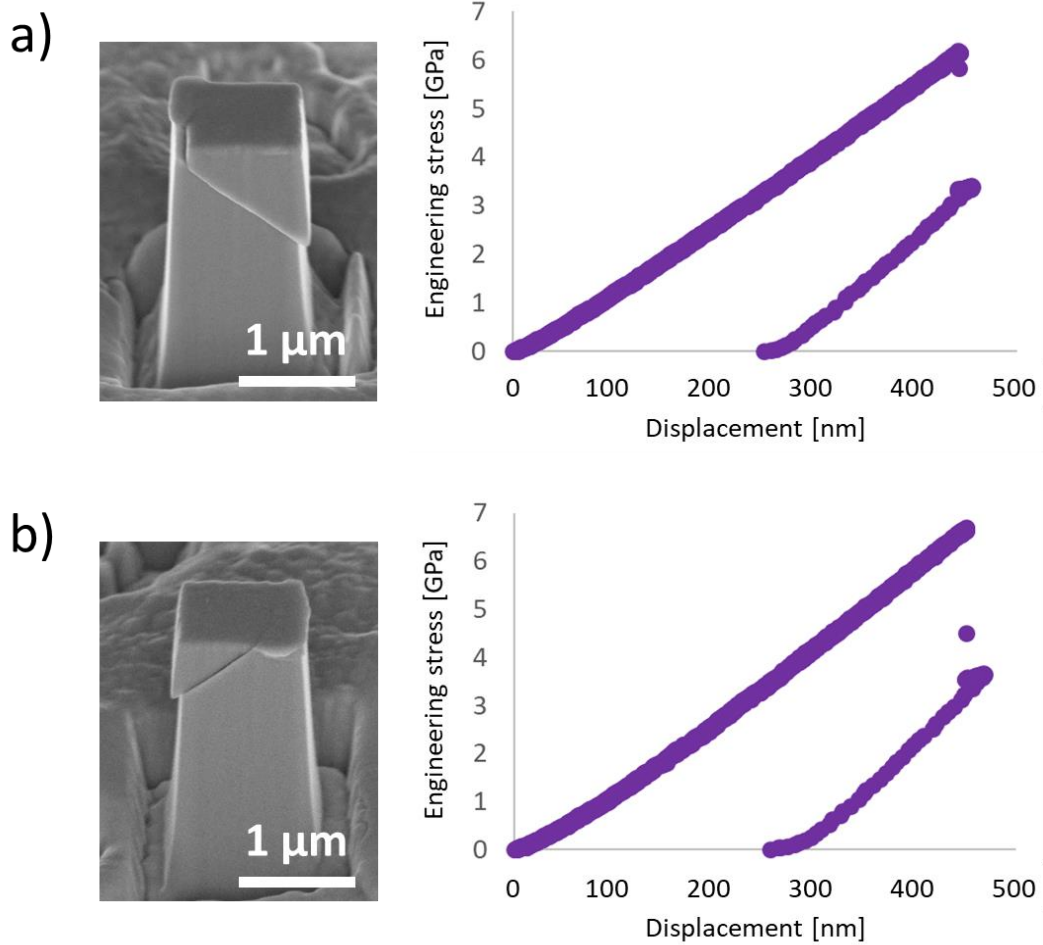

Figure S1 Post-deformation SEM images and engineering stress-indenter displacement curves for <111>-oriented (a) as-received and (b) hydrogenated micropillars with 1 µm mid-height width and 2.5 µm height.

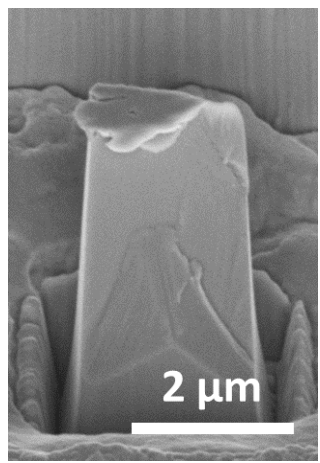

Figure S2 Post-deformation SEM image of a fractured micropillar (hydrogenated, <011>-oriented).

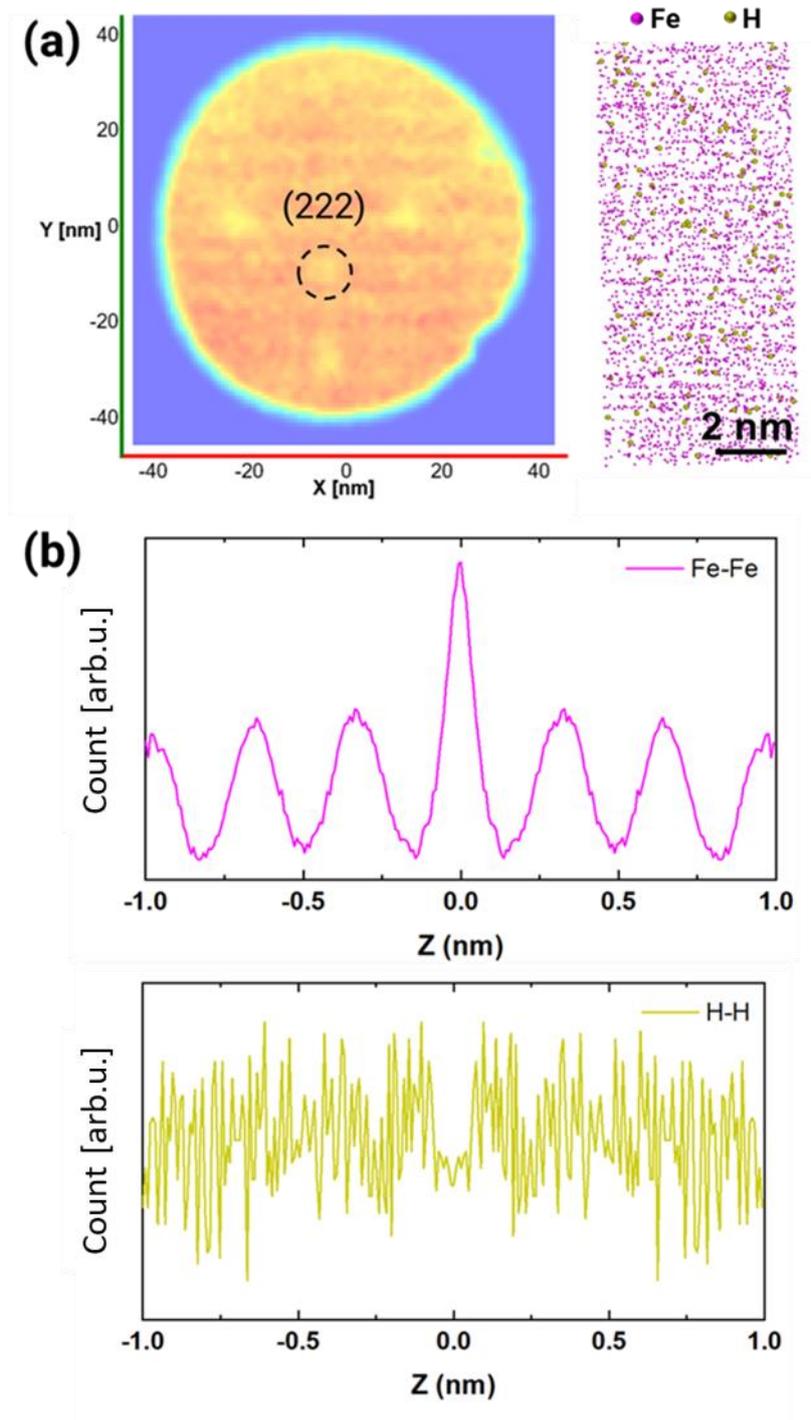

Figure S3 APT analysis of the as-received sample: (a) 2D density map of the APT volume taken from a grain oriented near [111] zone axis. Region of interest (ROI) is centred within the pole and the resolved {222} planes are shown next to the 2D density map. (b) z-Spatial distribution maps (SDMs) of Fe and H corresponding to the {222} planes in (a). Arb.u. denotes arbitrary unit.

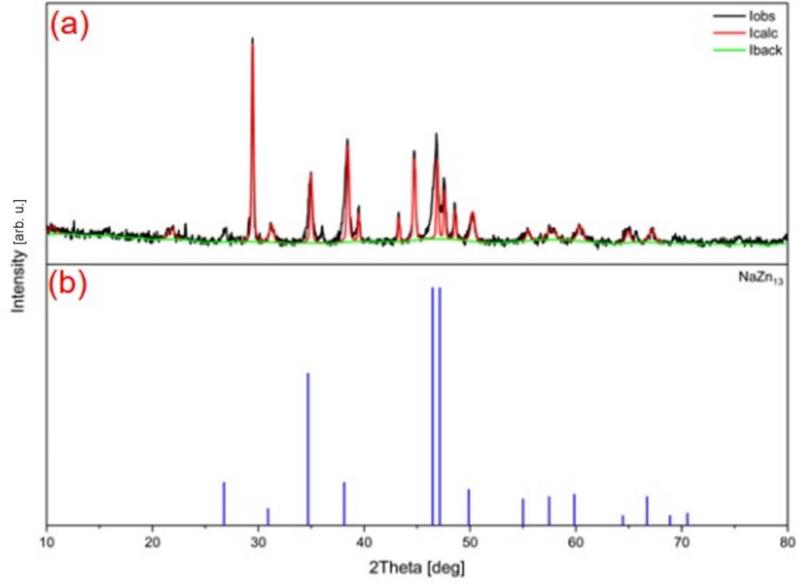

Figure S4  X-ray diffraction patterns of (a) the original alloy sample and (b) the reference of NaZn$_{13}$-type structure. The cell parameter of the Fm-3c type 1:13 phase was calculated to be a=11.58 Å and the extra peaks are from the α-Fe and La-rich phases. Arb.u. denotes arbitary unit.

Table S1  Chemical compositions of the 1:13 phase and the α-Fe phase in the as-received sample [at.%].

|  | Fe | La | Si | Mn | H |
|---|---|---|---|---|---|
| Matrix | 76.2 ± 0.6 | 7.4 ± 0.1 | 8.9 ± 0.1 | 2.1 ± 0.01 | 5.2 ± 0.8 |
| α-Fe | 91.3 ± 0.010 | -- | 1.5 ± 0.005 | 3.1 ± 0.006 | 3.8 ± 0.006 |

Table S2  Chemical compositions of the 1:13 phase and the α-Fe phase in the hydrogenated sample [at.%].

|  | Fe | La | Si | Mn | H |
|---|---|---|---|---|---|
| Matrix | 69.8 ± 1.2 | 6.8 ± 0.2 | 8.2 ± 0.2 | 1.9 ± 0.03 | 13.1 ± 1.5 |
| α-Fe | 92.0 ± 0.013 | -- | 1.5 ± 0.006 | 3.0 ± 0.009 | 2.9 ± 0.008 |